# Systematic study of E2 matrix elements in the framework of the Triaxial Projected Shell Model

*Stefan* Frauendorf[1*], *Syed* Rouoof[7], *Gowhar* Bhat[2], *Sheikh* Jehangir[3], *Nazira* Nazir[4,5,6], *Kouser* Qureshie[7], *Niyaz* Rather[7], and *Javid* Sheikh[6**]

[1]Department of Physics, University of Notre Dame, Notre Dame, Indiana 46556, USA
[2]Department of Physics, Government Degree College Shopian, Jammu and Kashmir, 192 303, India
[3]Department of Physics, Government Degree College Kulgam, Jammu and Kashmir, 192 231, India
[4]Institute of Physics (IOP), Sachivalya Marg, Bhubaneswar, 751005, India
[5]Homi Bhabha National Institute, Training School Complex, Anushakti Nagar, Mumbai, 400 094, India
[6]Department of Physics, University of Kashmir, Srinagar, 190 006, India
[7]Department of Physics, Islamic University of Science and Technology, Awantipora, 192 122, India

**Abstract.** Nuclides for which extended sets of E2 matrix elements have been measured by means of COULEX experiments are studied in the framework of the triaxial projected shell model (TPSM). The studies encompass: $^{70,72,74,76}$Ge, $^{76,78,80,82}$Se, $^{100}$Mo, $^{104}$Ru, $^{110}$Pd, $^{168}$Er, $^{186,188,190}$Os, $^{184}$Pt. The experimental energies of the ground band, of the quasi $\gamma$ band and of some excited $0^+$bands as well as their individual intra and inter band matrix E2 matrix elements are systematically accounted for by the microscopic TPSM calculations. The studied nuclei represent axial, rigid triaxial and soft triaxial shapes from the perspective of the collective model. The discriminating features are demonstrated by comparing the Kumar-Cline shape invariants derived from the TPSM results with the experimental ones. The TPSM uses the static triaxial deformation of the mean field as the only input adjusted to the experiment. The relation between the model's capability to account for the different shape characteristics and the superposition of multi quasiparticle excitations will be discussed.

## 1 Introduction

The concept of the nuclear shape has been extensively exposed in textbooks, for example [1], as the basis of the Collective Model (CM). The body fixed quadrupole moments define the parameters $\beta$ of deviation from spherical and $\gamma$ from axial shape. In the CM these are considered as dynamical degrees of freedom of the Bohr Hamitonian. From this perspective it has become custom to characterize nuclei as "rigid triaxial, $\gamma$ rigid" when $\gamma$ is restricted to a small range and "$\gamma$ soft" when the wave function is distributed over a wide range. There are numerous approaches that put the CM concept on a microscopic foundation called nowadays 5-Dimensional Bohr Hamiltonian (5DBH), which have been recently reviewed in Ref. [2].
Experimental information about the nuclear shape can be obtained by measuring the electric quadrupole (E2) moments, static and transitional, of the low-lying excitations by means of COULEX experiments and decay lifetimes. The poster studies several examples.
The Triaxial Projected Shell Model [3] is an alternative approach to describe the collective quadrupole excitations, which is employed in this poster. Since the model combines symmetry restoration with configuration mixing, it has become a powerful tool for a unified description of rotational bands, γ vibrations, quasiparticle excitations, chiral doublet bands, wobbling motion, and other spectroscopic phenomena in medium- and heavy-mass nuclei [4–10].
.

## 2 The Triaxial Projected Shell Model

The Triaxial Projected Shell Model (TPSM) is a microscopic shell-model framework that has been developed specifically for the analysis of collective excitations in deformed and transitional nuclei. It employs a basis constructed from triaxially deformed quasiparticle states projected on good angular momentum
The first step in a TPSM calculation is the construction of a triaxially deformed quasiparticle basis by diagonalizing the Nilsson Hamiltonian and incorporating pairing correlations by means of the BCS approximation. The triaxial Nilsson Hamiltonian is given by

$$\hat{H}_N = \hat{H}_0 - \frac{2}{3}\hbar\omega\left[\varepsilon\hat{Q}_0 + \varepsilon'\frac{\hat{Q}_{+2}+\hat{Q}_{-2}}{\sqrt{2}}\right], \quad (1)$$

where $\hat{H}_0$ represents the spherical single-particle Hamiltonian. The quantities $\varepsilon$ and $\varepsilon'$ characterize the axial and non-axial quadrupole deformations, respectively. The degree of triaxiality is given by

$$\gamma = tan^{-1}\left(\frac{\varepsilon'}{\varepsilon}\right). \quad (2)$$

The axial deformation is chosen from the experimental $B(E2, 2_1^+ \to 0_1^+)$ value or from self-consistent mean-field calculations, whereas the non-axial deformation is adjusted so that the calculated γ band head reproduces the observed excitation energy.

*Corresponding author: Stefan.G.Frauendorf.1@nd.edu
**Corresponding author: sjaphysics@gmail.com

The rotational symmetry is restored by the three-dimensional angular-momentum projection operator,

$$\hat{P}^{I}_{MK} = \frac{2I+1}{8\pi^2}\int d\Omega\, D^{I}_{MK}(\Omega)\hat{R}(\Omega), \quad (3)$$

where $D^{I}_{MK}(\Omega)$ denotes the Wigner rotation matrix and $\hat{R}(\Omega)$ represents the rotation operator defined by the Euler angles $\Omega = (\alpha, \beta, \gamma)$.

For even-even nuclei, the TPSM model space consists of the triaxial projected quasiparticle vacuum together with two-proton, two-neutron and four-quasiparticle configurations,

$$\hat{P}^{I}_{MK}|\Phi\rangle,\ \hat{P}^{I}_{MK}a^{\dagger}_{\pi1}a^{\dagger}_{\pi2}|\Phi\rangle,$$
$$\hat{P}^{I}_{MK}a^{\dagger}_{\nu1}a^{\dagger}_{\nu2}|\Phi\rangle,\ \hat{P}^{I}_{MK}a^{\dagger}_{\pi1}a^{\dagger}_{\pi2}a^{\dagger}_{\nu1}a^{\dagger}_{\nu2}|\Phi\rangle, \quad (4)$$

where $\Phi\rangle$ denotes the triaxially deformed quasiparticle vacuum and $a^{\dagger}$ is the the quasiparticle creation operator. The inclusion of multiquasiparticle configurations enables TPSM to describe both collective rotational structures and quasiparticle excitations within a unified microscopic framework.

The projected basis states are used to diagonalize the pairing-plus-quadrupole shell-model Hamiltonian,

$$\hat{H} = \hat{H}_0 - \frac{1}{2}\chi\sum_{\mu}\hat{Q}^{\dagger}_{\mu}\hat{Q}_{\mu} - G_M\hat{P}^{\dagger}\hat{P} - G_Q\sum_{\mu}\hat{P}^{\dagger}_{\mu}\hat{P}_{\mu}, \quad (5)$$

where $\chi$ is the quadrupole-quadrupole interaction strength, while $G_M$ and $G_Q$ denote the monopole and quadrupole pairing strengths, respectively. The interaction strengths are chosen to obey selfconsistency with the empirical pairing gaps and the adopted deformation parameters, which are the input of the TPSM.

The eigenfunctions, obtained after configuration mixing, are expressed as

$$|\Psi^{\sigma}_{IM}\rangle = \sum_{\kappa,K} f^{\sigma I}_{\kappa K}\, \hat{P}^{I}_{MK}|\Phi_{\kappa}\rangle, \quad (6)$$

where $f^{\sigma I}_{\kappa K}$ are the mixing amplitudes and $\sigma$ distinguishes different states having the same angular momentum.

The reduced E2 matrix elements are obtained from the projected many-body wave functions according to

$$\left\langle \Psi^{\sigma_f}_{I_f} \left\| \hat{E}_2 \right\| \Psi^{\sigma_i}_{I_i} \right\rangle =$$
$$\sum_{\kappa_f,\kappa_i}\sum_{K_f,K_i} f^{\sigma_f I_f}_{\kappa_f K_f} f^{\sigma_i I_i}_{\kappa_i K_i} \left\langle \Phi_{\kappa_f} \left| \hat{P}^{I_f}_{K_f}\hat{E}_{2\mu}\hat{P}^{I_i}_{K_i} \right| \Phi_{\kappa_i} \right\rangle \quad (8)$$

The calculations employ effective charges of $e_{\pi} = 1.5e$ for protons and $e_{\nu} = 0.5e$ for neutrons to account for core-polarization effects.

## 3 Comparison with experiment

The experimental energies and E2 matrix for $^{104}$Ru, $^{72,76}$Ge $^{168}$Er, $^{186,188,190}$Os, $^{184}$Pt have been compared with calculated ones in Refs. [8,9]. Figs. 1-6, 9,10 show the nuclides studied in Ref. [9] as examples. The agreement with the experiment is remarkable. The TPSM accounts for the not shown nuclei studied in Refs. [10] with similar accuracy. The selection encompasses nuclei that in the context of the 5DCBH are classified as “axial” ($^{168}$Er), “soft triaxial, $\boldsymbol{\gamma}$ soft” ($^{72}$Ge) and “rigid triaxial, $\boldsymbol{\gamma}$ rigid” ($^{192}$Os) for which the TPSM reproduces the experimental energies and matrix elements equally well. A list of classifying quantities is given in Ref. [5]. One of them is the staggering parameter

$$S(I) = (E(I) - 2E(I-1) + E(I-2))/E(2^+_1) \quad (9)$$

of the energies of the $\boldsymbol{\gamma}$ band. Even-$I$-down indicates $\boldsymbol{\gamma}$ soft, odd-I-down indicates $\boldsymbol{\gamma}$ rigid.

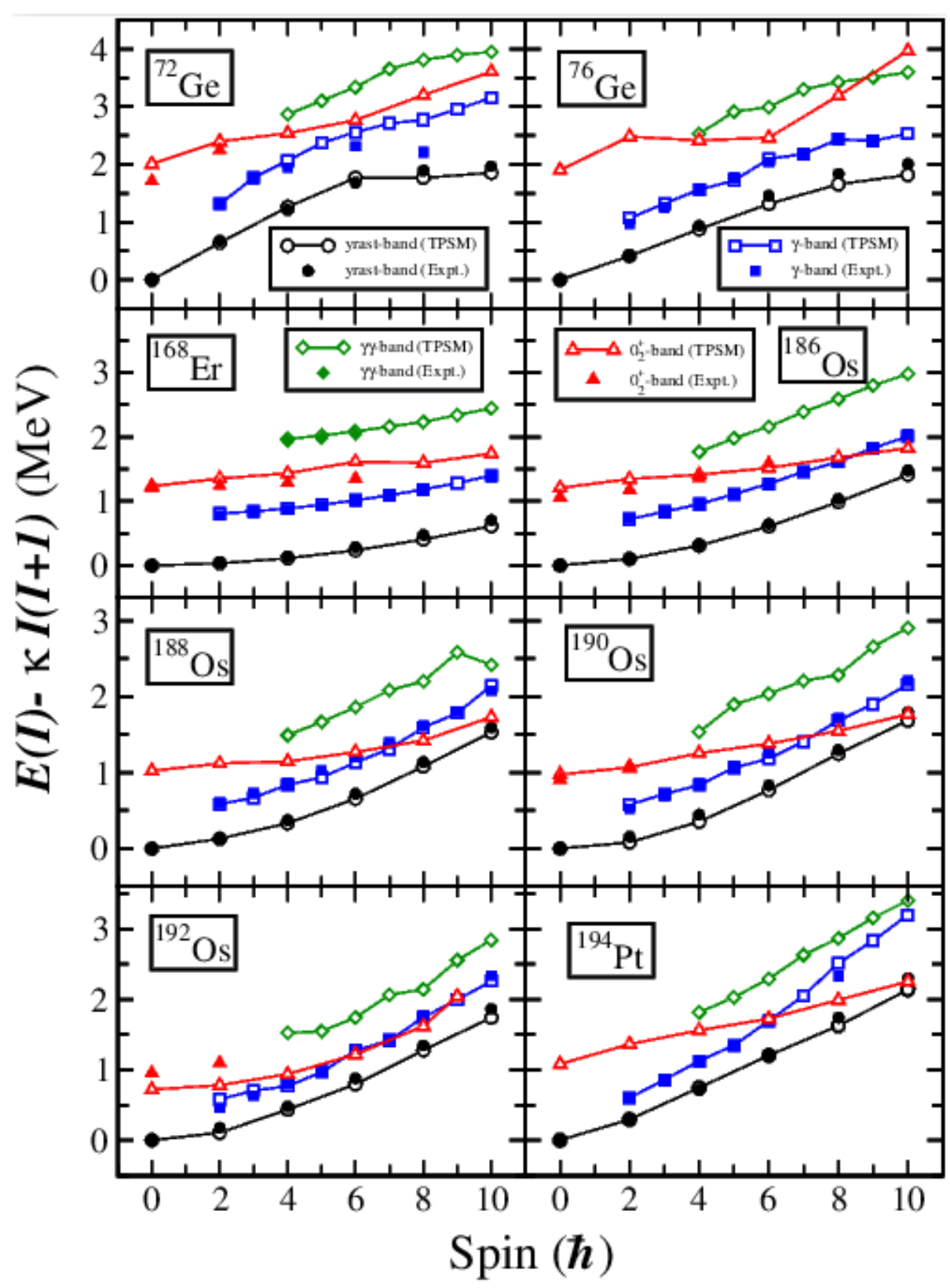

**Fig. 1.** TPSM and Experimental energies. The scaling factor $\kappa = 32.32^{-5/3}$. From Ref. [9]

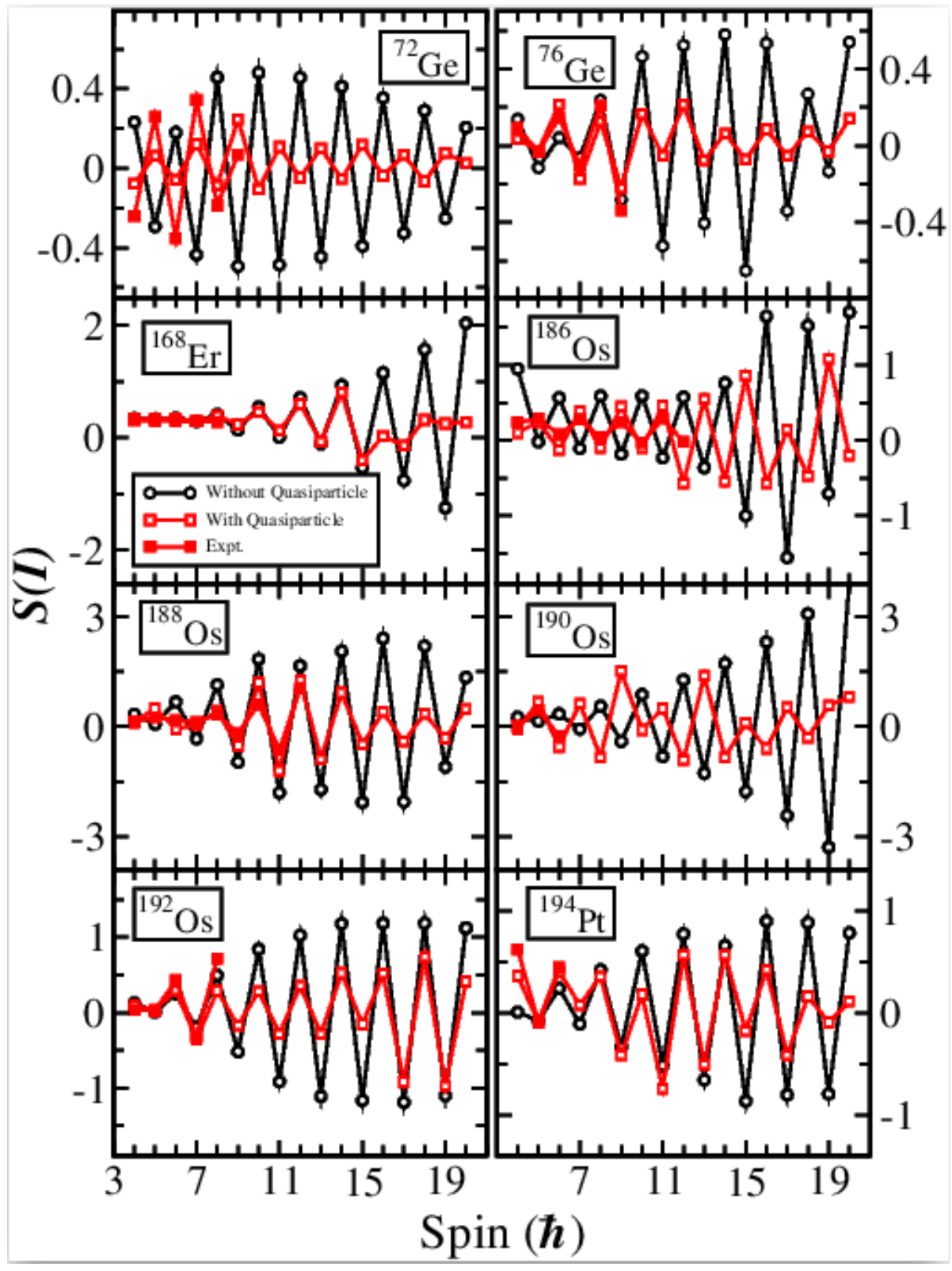

**Fig. 2.** Staggering parameter *S(I)* of the $\gamma$-band. The black open circles show the TPSM excluding the quasiparticle excitations and the red open circle the TPSM including them. The red squares display the experiment. From Ref. [9].

As seen in Fig. 2, if the TPSM includes only the vacuum configuration it gives only the odd-*I*-down pattern of a $\boldsymbol{\gamma}$ rigid nucleus, which is expected because the TPSM assumes a triaxial mean field. Only the admixture of the quasiparticle excitations generates the even-*I*-down feature of $\boldsymbol{\gamma}$ softness.

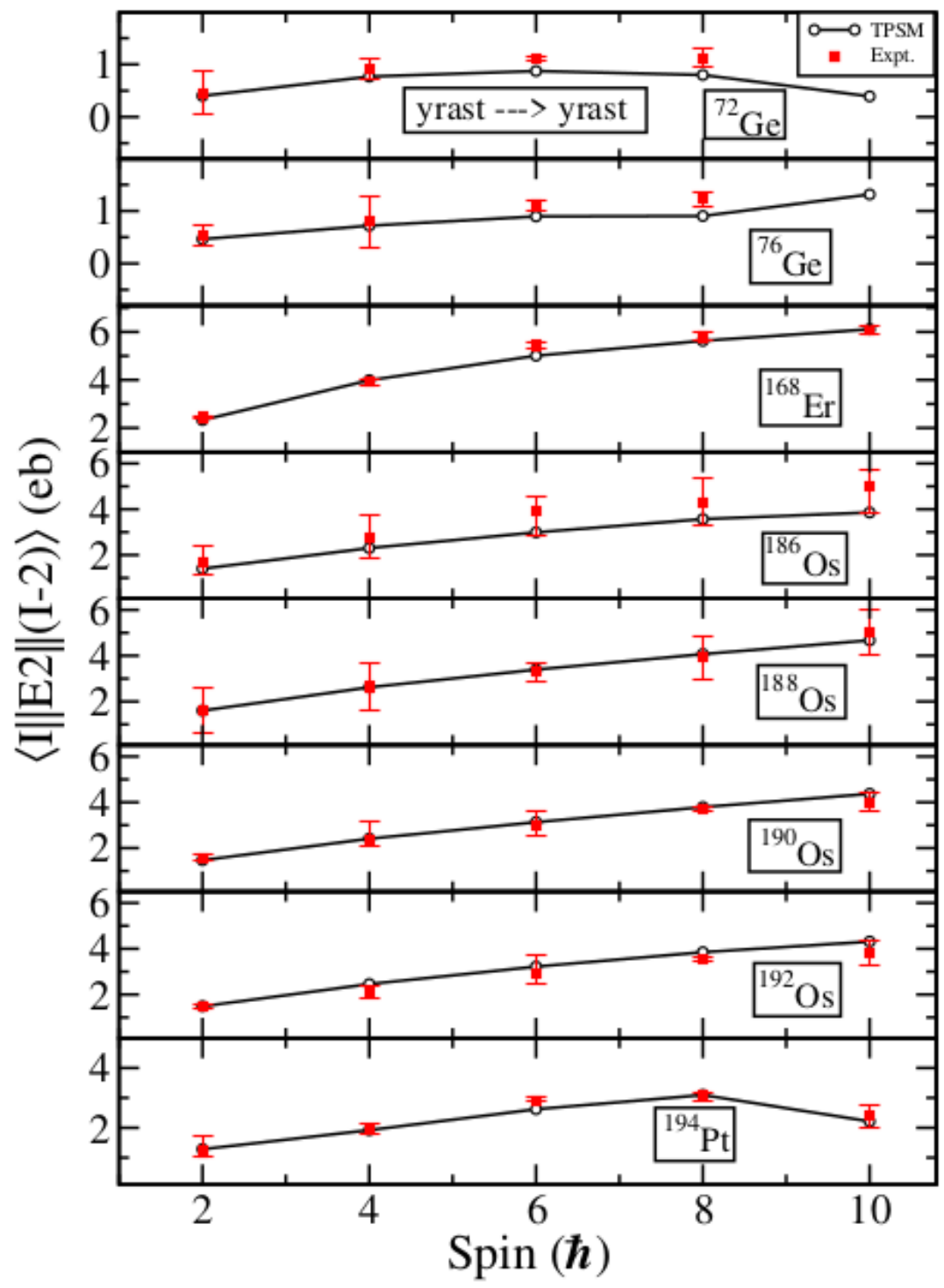

**Fig. 3.** Reduced intra band *E2* matrix elements for transitions yrast to yrast. From Ref. [9].

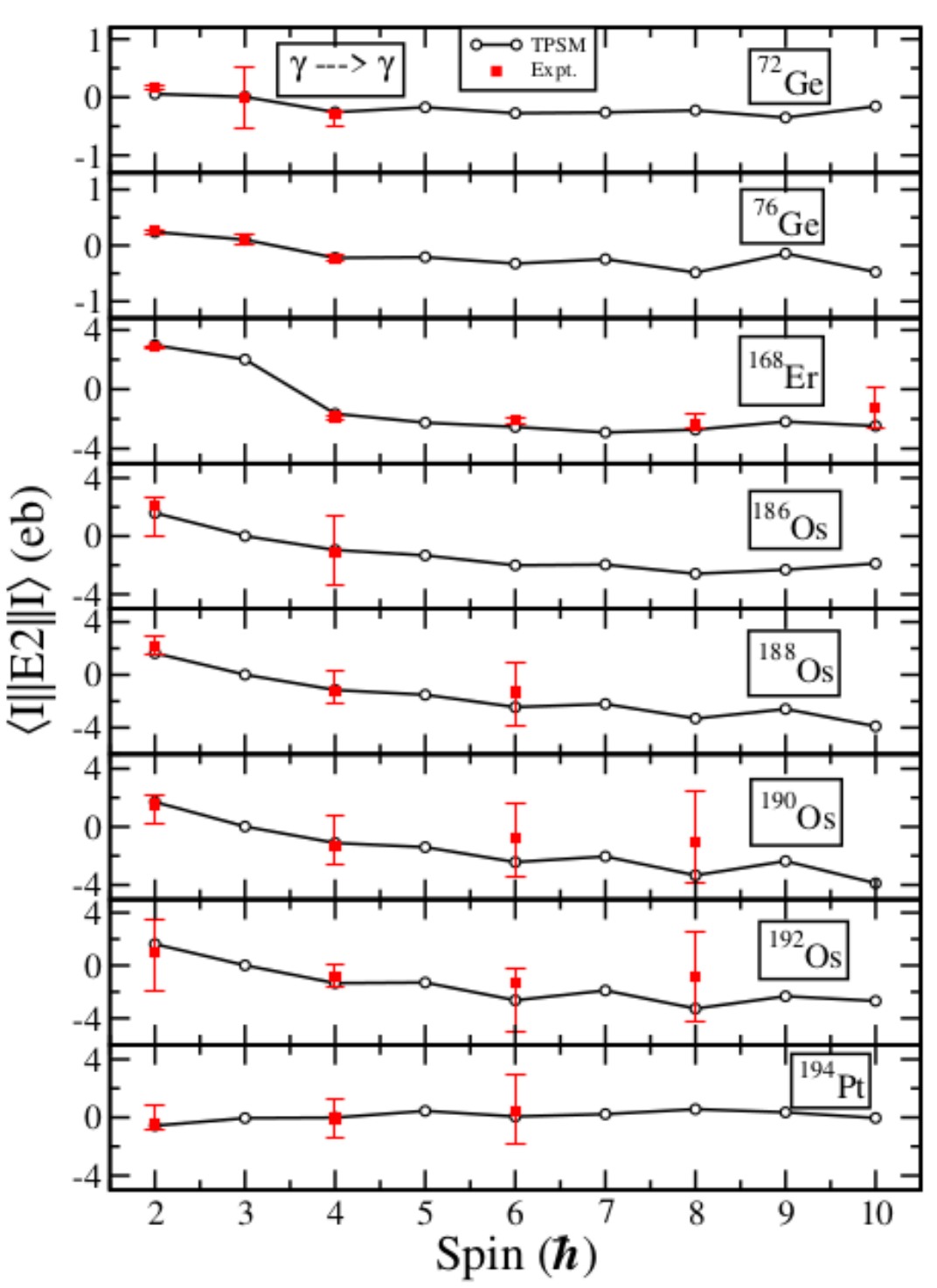

**Fig. 4.** Reduced diagonal *E2* matrix elements for transitions $\gamma$ band to $\gamma$ band, which represent the static quadrupole moments. From Ref. [9].

The staggering parameters *S(I)* and the $B(E2)$ values within and between the yrast and $\boldsymbol{\gamma}$ bands of a number of additional nuclei have been studied in Ref. [5]. The results are similar. The TPSM agrees very well the data and the admixture of the quasiparticle excitations accounts for $\boldsymbol{\gamma}$ softness.

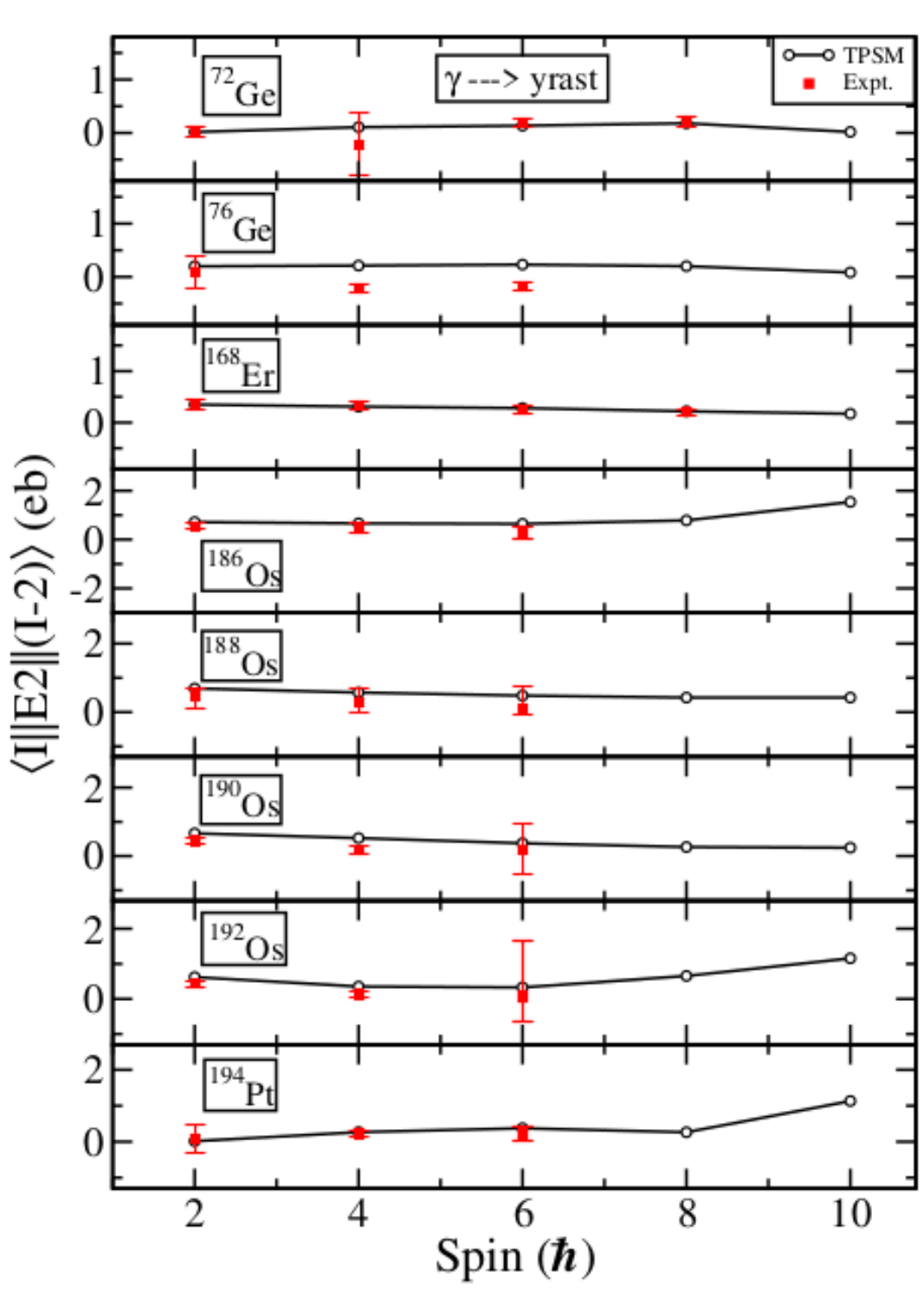

**Fig. 5**. Reduced inter band *E2* matrix elements for transitions $\gamma$ band to yrast band. From Ref. [9].

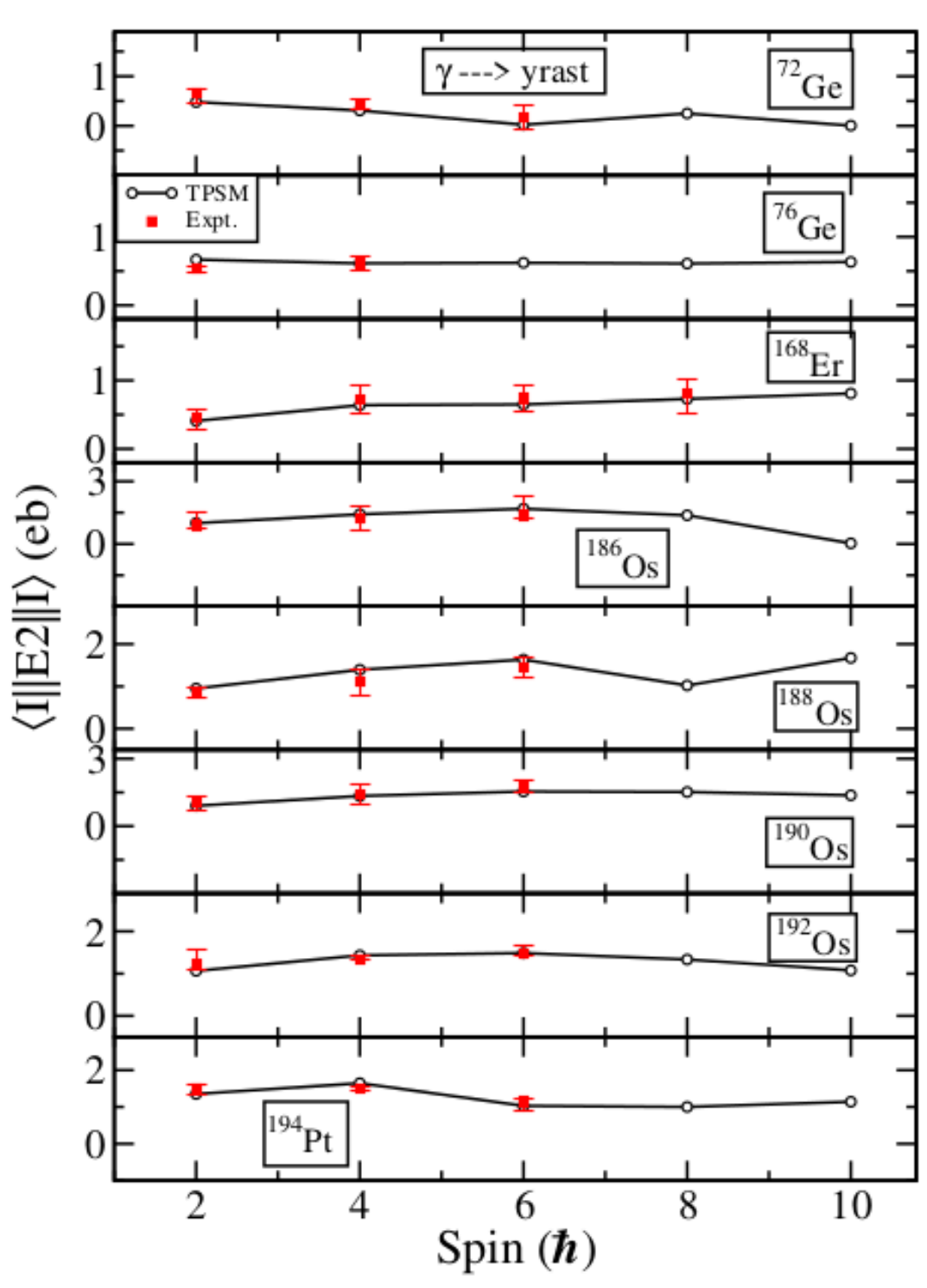

**Fig. 6.** Reduced inter band *E2* matrix elements for transitions the $\gamma$ band to the yrast band. From Ref. [9].

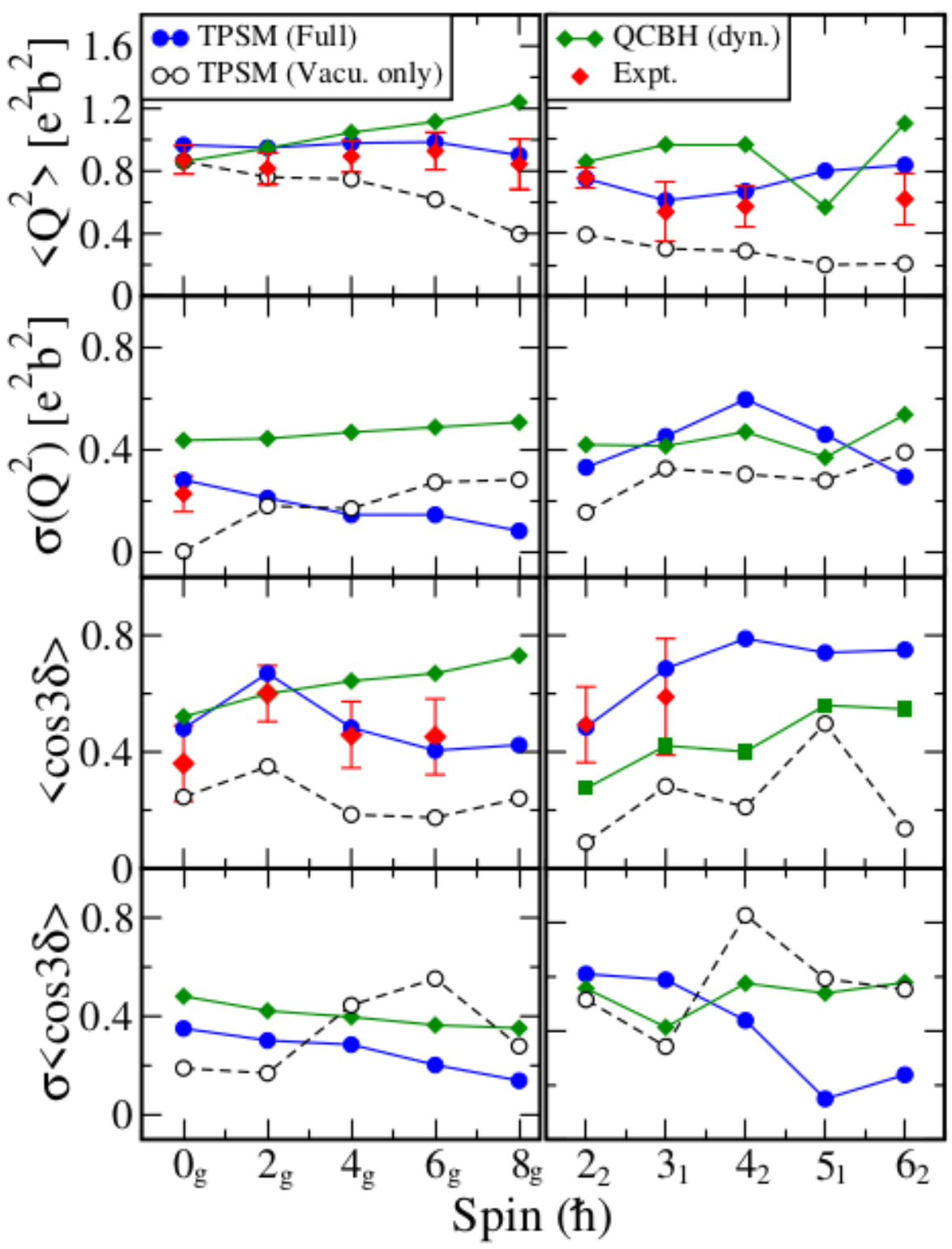

**Fig. 7.** TPSM, QCBH, and experimental shape invariants for yrast and $\gamma$ bands in $^{104}$Ru. From Ref. [8].

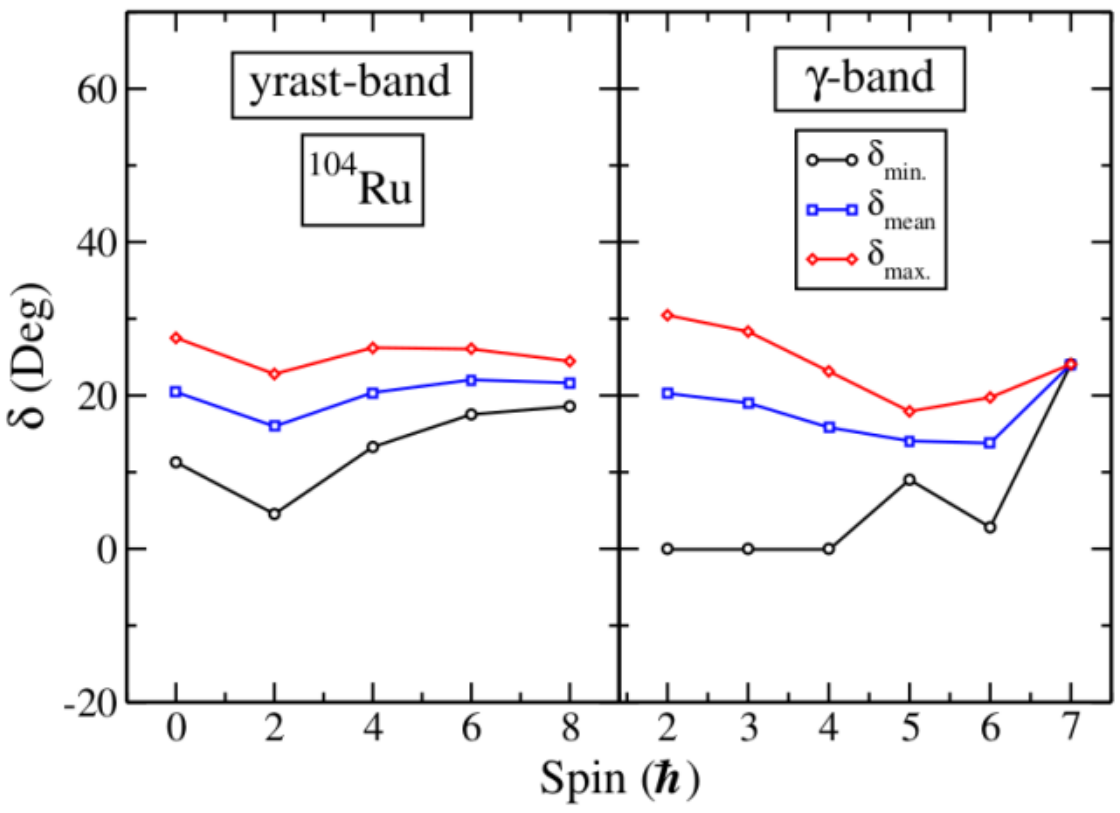

**Fig. 8.** Bands of the triaxiality parameter $\delta$, assuming normal distributions, with the dispersions $\sigma(\cos 3\delta)$ and the centroids $\langle\cos 3\delta\rangle$. Derived from the calculations in Fig. 7.

## 4 Quadrupole Shape Invariants

The TPSM reduced electric-quadrupole matrix elements provide a very good description of the collective quadrupole response of the nuclear states. Instead of comparing individual *E2* transition and classifying the quadrupole mode by some characteristic elements (see Ref. [5]) one can refer to the quadrupole shape invariants introduced by Kumar [11], which provide direct information about the mean values of the deformation and triaxiality parameters of an individual state as well as about their fluctuations.

For a state $|S\rangle$ with angular momentum *S*, the second-order quadrupole invariant is obtained by coupling two *E2* tensors to total angular momentum zero. In terms of the reduced *E2* matrix elements it is

$$\langle Q^2\rangle = \frac{1}{2S+1}\sum_R |\langle S\|E2\|R\rangle|^2 \tag{10}$$

and measures the mean intrinsic quadrupole moment. The third order invariant

$$\langle Q^3\cos 3\delta\rangle = \mp\sqrt{\frac{35}{2}}\frac{1}{2S+1}\sum_{RT}\langle S\|E2\|R\rangle\langle R\|E2\|T\rangle\langle T\|E2\|S\rangle \times \begin{Bmatrix} 2 & 2 & 2\\ S & T & R\end{Bmatrix} \tag{11}$$

contains the information about the triaxiality parameter $\delta \approx \gamma$. In contrast to the quadratic invariant, the cubic invariant depends on the relative phases of the reduced *E2* matrix. elements A dimensionless measure of the triaxiality parameter is defined as

$$\langle\cos 3\delta\rangle = \frac{\langle Q^3\cos 3\delta\rangle}{\langle Q^2\rangle^{3/2}}, \tag{12}$$

and the mean value of the triaxiality parameter as $\langle\delta\rangle = \arccos[\langle\cos 3\delta\rangle]/3$.

In order to quantify fluctuations of the shape, the dispersion of the quadrupole magnitude

$$\sigma(Q^2) = \sqrt{\langle Q^4\rangle. - \langle Q^2\rangle^2}, \tag{13}$$

is introduced, which requires evaluating the fourth-order invariant $\langle Q^4\rangle$. Analogously, the fluctuations in triaxiality are obtained from the dispersion

$$\sigma(\cos 3\delta) = \left[\frac{\langle(Q^3\cos 3\delta)^2\rangle}{\langle Q^6\rangle} - \left(\frac{\langle Q^3\cos 3\delta\rangle}{\langle Q^2\rangle^{3/2}}\right)\right], \tag{14}$$

which requires the evaluation of $\langle(Q^3\cos 3\delta)^2\rangle$ and $\langle Q^6\rangle$. Expressions for the higher order of shape invariants are given in Ref. [12], which was used to evaluate them from the TPSM and experimental E2 matrix elements.

Fig. 7 compares the quadrupole invariants calculated from the TPSM matrix elements with the ones obtained from the COULEX matrix elements for $^{104}$Ru. The TPSM very well describes the mean values of the deformation and the triaxiality for the states of the yrast and $\gamma$ bands. The magnitude of the deformation changes weakly with *I.* Both bands are triaxial on the prolate side, $\langle\delta\rangle \approx 20°$. The TPSM calculation without the quasiparticle admixtures substantially deviates from the experiment, which demonstrates their importance.

The figure also displays calculation of the 5DBH class which is called QCBH (see Ref. [8] for details). The TPSM better agrees with the experiment.

The number of COULEX matrix elements is too small to calculate the dispersions for the states but one, which indicates a stable ground state deformation.

The calculated dispersions may be used to classify the quadrupole mode. Assuming a normal distribution for $\cos 3\delta$, one can define a lower bound and an upper bound for $\delta$ within which 70 % of the probability is localized. Fig. 8 shows the limits together with the mean value. The wide probability band classifies $^{104}$Ru as "$\gamma$ soft" with a preference for prolate shape, which stabilizes with increasing *I*.

Fig. 9 compares the TPSM centroids of the triaxiality parameter $\langle\cos 3\delta\rangle$ for the yrast and $\gamma$ bands of the nuclides studied in Ref. [9] with their experimental values. Again, the agreement is remarkable. Except the prolate $^{168}$Er, most cases show substantial triaxility with a preference ofprolate shape. Only $^{194}$Pt tends to oblate shape.

Fig. 10 displays the probability stripes calculated from the centroids in Fig. 9 and the dispersions published in Ref. [9].

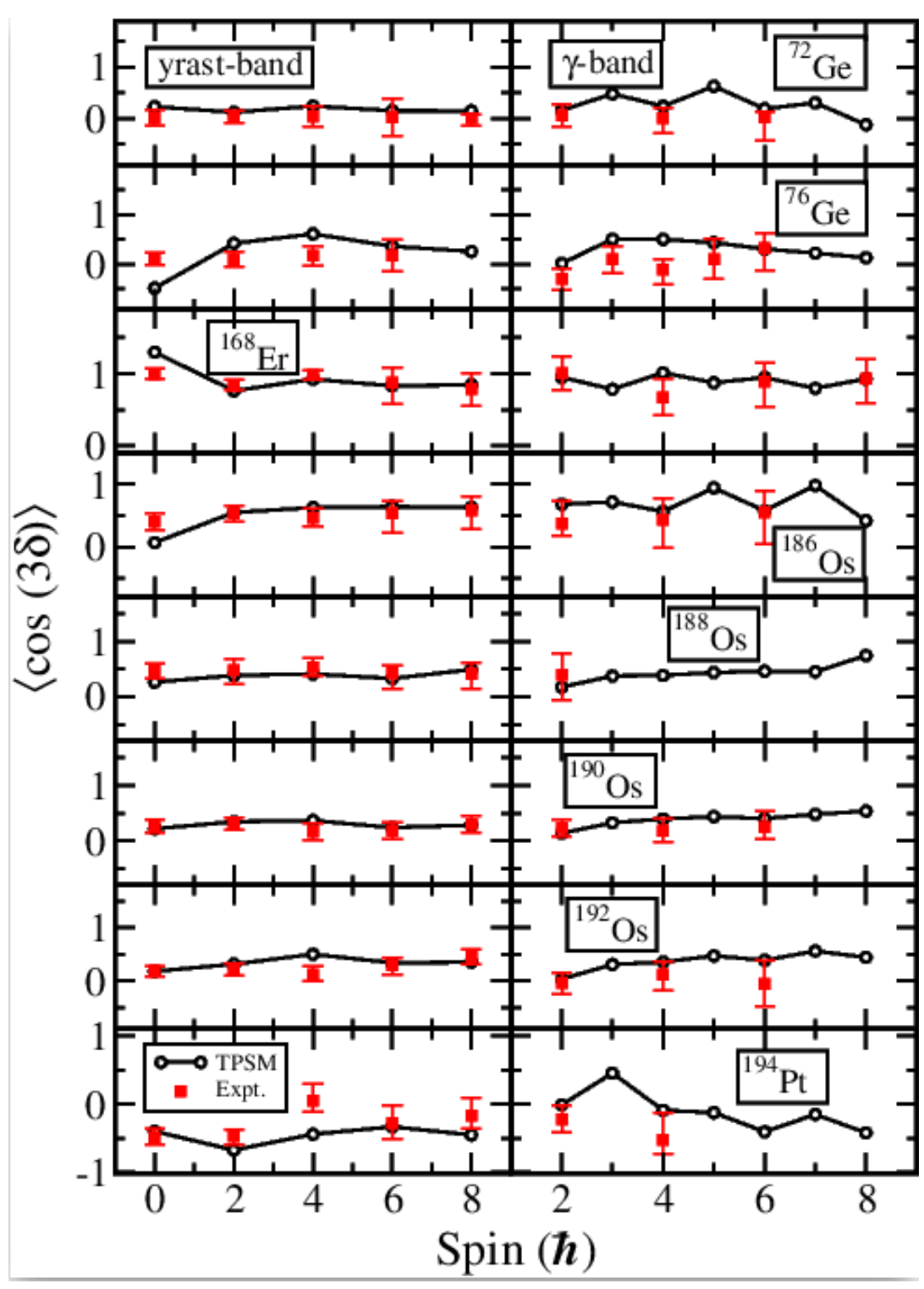

**Fig. 9.** Centroid $\langle\cos 3\delta\rangle$ of the yrast and $\gamma$ bands studied in Ref. [9]. From Ref. [9]

The prolate shape of $^{168}$Er is rigid. The wide probability stripes of $^{72,76}$Ge classify the yrast bands of these nuclei as $\gamma$ soft, where $^{72,}$Ge becomes rigid triaxial with *I*. The $\gamma$ bands of the two isotopes are rigid triaxial. The yrast band of $^{194}$Pt is $\gamma$ soft. Its $\gamma$ band staggers between $\gamma$ softness for even *I* and rigidity for odd *I*. The reason of it is unclear.
All Os isotopes are rigid triaxial according to the shape invariants obtained from the TPSM E2 matrix elements.
As seen in Fig, 2, the staggering pattern of the TPSM $\gamma$ band energies are: even-*I*-down, -up, -down, -up, for *A*=186, 188, 190, 192, respectively, which agrees with experimental $S(I)$ values. In the context of the phenomenological 5DBH (see Ref. [2]), the change of the $S(I)$ pattern with $A$ indicates a back and forth between $\gamma$ soft and $\gamma$ rigid, which contradicts Fig. 10 which indicates that all Os isotopes are $\gamma$ rigid.
The apparent contradiction reveals a limitation of the CM. As discussed in Ref. [5,6], the staggering reflects the repulsion between the even-*I* states of the $\gamma$ band, the ground band below them and the $0^+$ bands lying above them. The odd-*I* states are not shifted because $0^+$ bands have only even-*I* members. For the phenomenological 5DBH the upper $0^+$ band represents the first collective vibrational excitation in the $\gamma$ degree of freedom. In $\gamma$ soft nuclei it has a low energy. Its repulsion is stronger than the repulsion from the ground band below and the even-*I* states of the $\gamma$ band are pushed below the odd-*I* ones. In $\gamma$ rigid nuclei it has a large energy. Its repulsion is weaker than the repulsion from the ground band below and the even-*I* states of the $\gamma$ band are pushed below the odd-*I* ones.
This correlation is the consequence of the simple collective nature of the upper $0^+$ band. In the case of the TPSM the eexcited $0^+$ bands have a more complex nature. They are superpositions of two-quasiparticle states and sensitive to the position of the single particle states with respect to the Fermi level, which may generate a different $A$ dependence of $S(I)$ than the direct information on $\gamma$ softness from the E2 matrix elements.

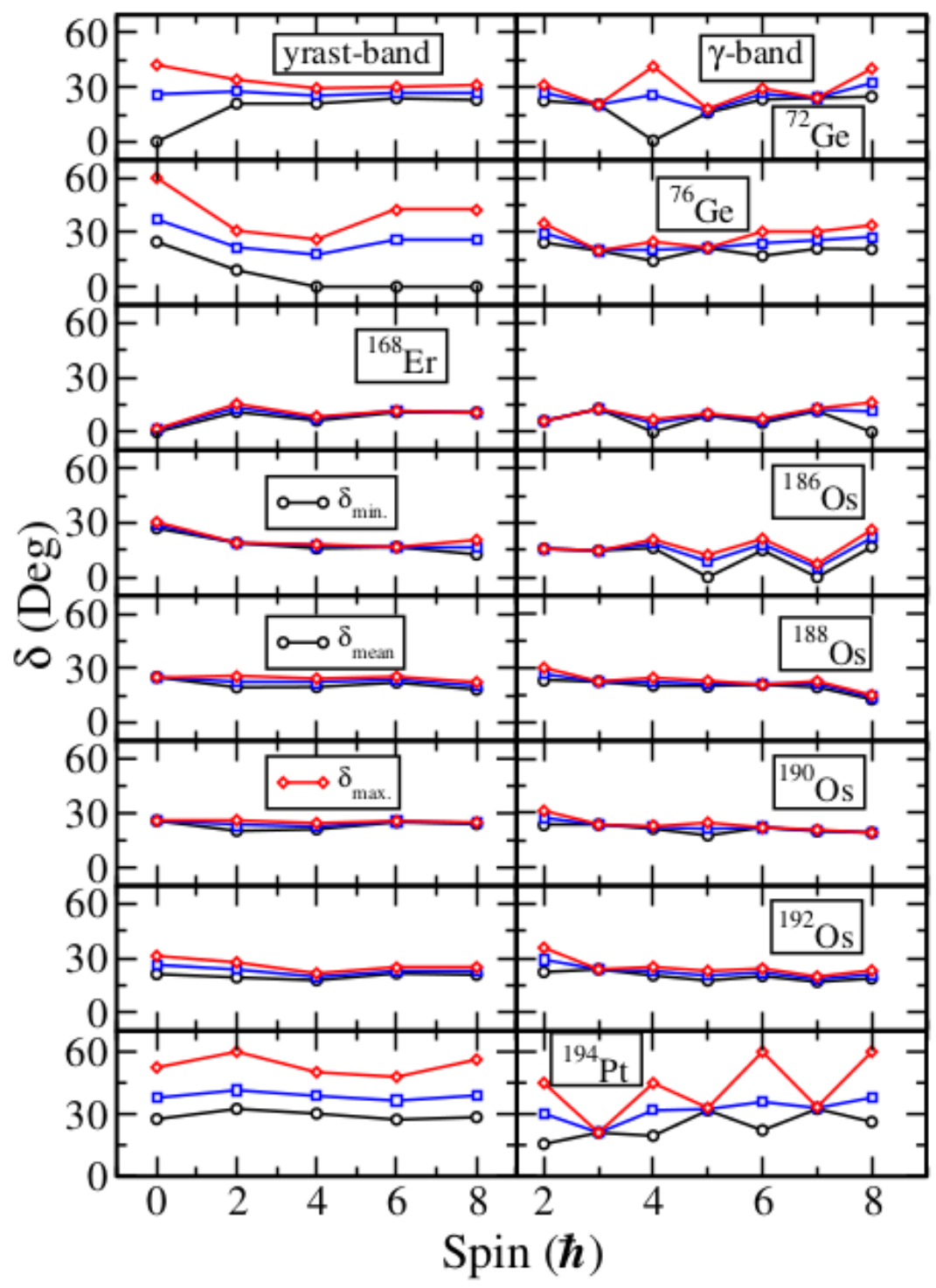

**Fig. 10.** Bands of the triaxiality parameter $\delta$, assuming normal distributions, with the dispersions $\sigma(\cos 3\delta)$ and the centroids $\langle\cos 3\delta\rangle$.

## 5 Summary and conclusions

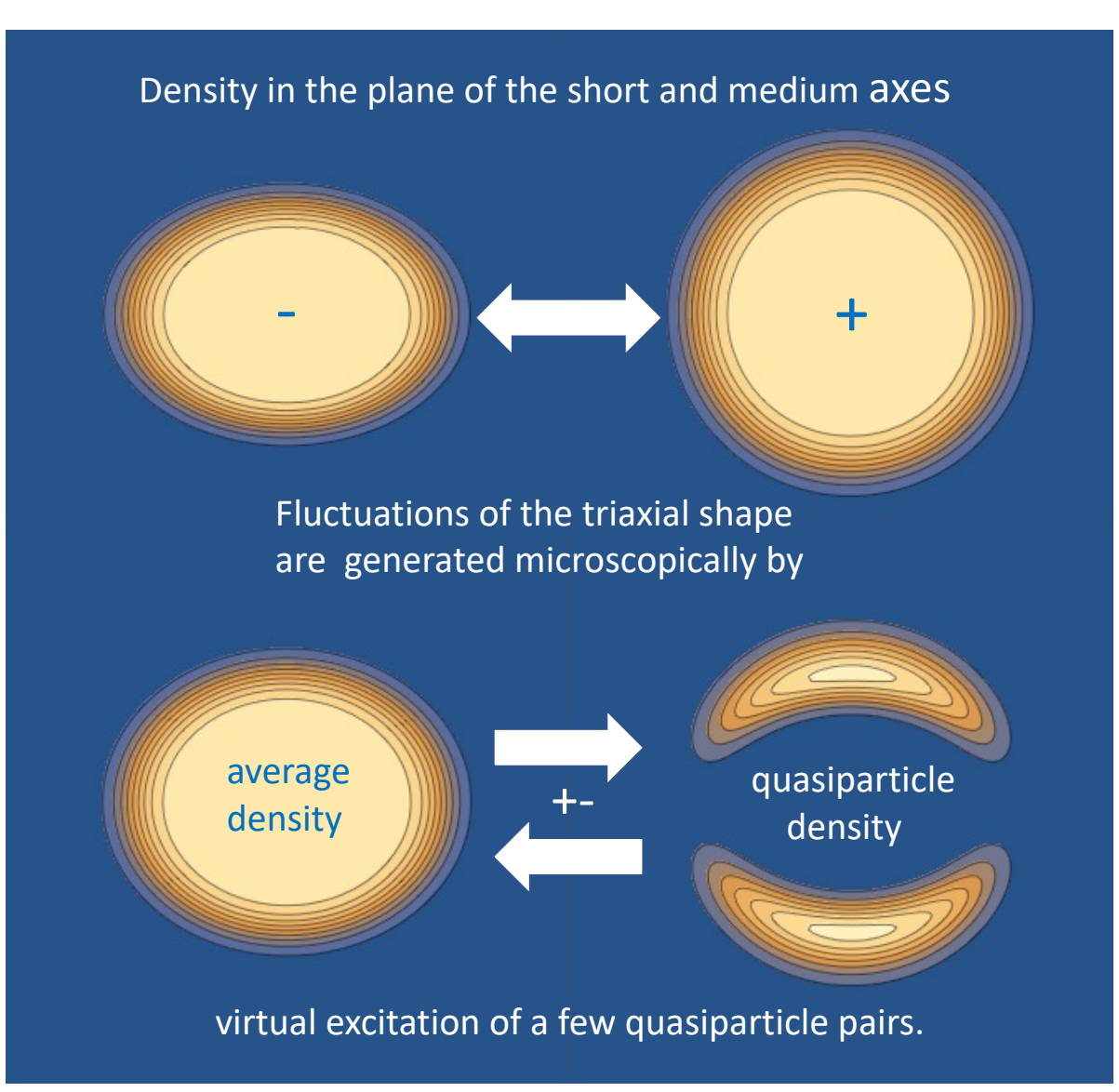

**Fig. 11** Illustration of shape fluctuations generated by quasiparticle admixtures

The nuclei $^{72,76}$Ge, $^{104}$Ru, $^{168}$Er, $^{186,188,190}$Os have been studied in the framework of the TPSM. The model systematically accounts for the experimental energies of the ground band, of the quasi $\gamma$ band as well as for the extended sets individual intra and inter band matrix E2 matrix elements from COULEX experiments. The new study of $^{70,74}$Ge, $^{76,78,80,82}$Se, $^{100}$Mo, $^{104}$Ru, $^{110}$Pd in Ref. [8], the results of which have not been shown, achieves a similar agreement between the TPSM and the COULEX E2 matrix elements as for the discussed examples.
The second, third, fourth and sixth order Kumar-Cline shape invariants have been calculated from the E2 matrix elements, which give the mean deformation and its fluctuations as well as the mean triaxiality and its fluctuations. This analysis leads to a direct classification of nuclei as being axial, triaxial $\gamma$ soft or $\gamma$ rigid.
The example of the Os isotopes demonstrates that the alternative classification based on the staggering of the $\gamma$ band energies is reflects a correlation that only holds in the framework of the phenomenological 5DBH but may be absent in the microscopic TPSM approach.

The fact that the TPSM very well describes nuclei with $\gamma$ soft characteristics may be appear surprising on first sight, because it starts from a mean field with a fixed mean triaxiality. Its success indicates that the configuration space of the pertaining angular momentum projected zero-, two- and four-quasiparticle is large enough to account for the fluctuations. Fig. 11 illustrates the mechanism. The upper panel shows the shape fluctuations as a collective phenomenon. The surface oscillates around a mean position where + and – correspond to the maximal elongation. The lower panel shows the microscopic picture. The vacuum represents the mean shape of the density distribution. The quasiparticle excitations have density distributions that deviate from the vacuum. Their admixture changes the shape total density. In a time-dependent formulation their contributions oscillate, i. e. the shape fluctuates. It remains to be seen to what extend the present version of thee TPSM can account for more drastic shape changes than encountered in the studied nuclei.

**References**

1. A. Bohr and B. Mottelson, Nuclear Structure II, W. A. Benjamin Inc. (1975)
2. S. Frauendorf, The low-energy quadrupole mode in nuclei, Int. J. of Modern. Phys. E 24, No. 9, 1541001 (2015), arxiv:150606287
3. J. A. Sheikh and K. Hara, Triaxial Projected Shell Model Approach. Phys. Rev. Lett. **82**, 3968 (1999).
4. J. A. Sheikh, G. H. Bhat, W. A. Dar, S. Jehangir, and P. A. Ganai, Microscopic nuclear structure models and methods: chiral symmetry, wobbling motion and $\gamma$–bands. Phys. Scr. **91**, 063015 (2016).
5. S. P. Rouoof, N. Nazir, S. Jehangir, G. H. Bhat, J. A. Sheikh, N. Rather, and S. Frauendorf, Fingerprints of the triaxial deformation from energies and $B(E2)$ transition probabilities of $\gamma$-bands in transitional and deformed nuclei. Eur. Phys. J. A **60**, 40 (2024).
6. S. Jehangir, G. H. Bhat, J. A. Sheikh, S. Frauendorf, W. Li, R. Palit, and N. Rather, Triaxial projected shell model study of $\gamma$–bands in atomic nuclei. Eur. Phys. J. A **57**, 308 (2021).
7. S. Jehangir, N. Nazir, G. H. Bhat, J. A. Sheikh, N. Rather, S. Chakraborty, and R. Palit, Extended triaxial projected shell model approach for odd-neutron nuclei. Phys. Rev. C **105**, 054310 (2022).
8. N. Nazir, S. Jehangir, S. P. Rouoof, G. H. Bhat, J. A. Sheikh, N. Rather, and S. Frauendorf, Microscopic aspects of $\gamma$ softness in atomic nuclei. Phys. Rev. C **107**, L021303 and Editor's Choice (2023)
9. S. P. Rouoof, N. Nazir, S. Jehangir, G. H. Bhat, J. A. Sheikh, N. Rather, and S. Frauendorf, Microscopic investigation of E2 matrix elements in atomic nuclei. Phys. Rev. C **111**, 054309 (2025).
10. S. P. Rouoof, N. Nazir, S. Jehangir, G. H. Bhat, J. A. Sheikh, N. Rather, and S. Frauendorf, Microscopic investigation of E2 matrix elements in atomic nuclei II. Phys. Rev. C, submitted (2026).
11. K. Kumar, Intrinsic Quadrupole Moments and Shapes of Nuclear Ground States and Excited States. Phys. Rev. Lett. **28**, 249 (1972)
12. D. Cline, T. Czosnyka, A. Hayes, P. Napiorkowski, N. Warr, and C. Wu, *Gosia User Manual for Simulation and Analysis of Coulomb Excitation Experiments* (University of Rochester, Rochester, NY, 2012).